\documentclass[aps,prl,10pt,twocolumn,tightenlines,superscriptaddress,floatfix]{revtex4}  
\usepackage{graphicx}
\usepackage[T1]{fontenc}
\usepackage[utf8]{inputenc} 
\usepackage{times} 
\usepackage{dcolumn}

\newcommand{\supp}{Supplemental Material \cite{prl-supp}}    
\newcommand{\cm}{cm$^{-1}$}
\newcommand{\comp}{Supplemental Material \cite{prl-supp}}

\begin{document}

\title{   Polyyne Electronic and Vibrational Properties under Environmental Interactions }

\author{  Marius Wanko }
\affiliation{ Nano-Bio Spectroscopy Group and European Theoretical Spectroscopy Facility (ETSF), Universidad del Pa\'{i}s Vasco, CFM CSIC-UPV/EHU-MPC \& DIPC, 20018 San Sebasti\'{a}n, Spain }

\author{  Seymur Cahangirov }
\affiliation{ Nano-Bio Spectroscopy Group and European Theoretical Spectroscopy Facility (ETSF), Universidad del Pa\'{i}s Vasco, CFM CSIC-UPV/EHU-MPC \& DIPC, 20018 San Sebasti\'{a}n, Spain }
\affiliation{ UNAM-National Nanotechnology Research Center, Bilkent University, 06800 Ankara, Turkey }

\author{  Lei Shi }
\affiliation{ University of Vienna, Faculty of Physics, 1090 Wien, Austria }

\author{  Philip Rohringer }
\affiliation{ University of Vienna, Faculty of Physics, 1090 Wien, Austria }

\author{  Zachary J. Lapin }
\affiliation{ ETH Z\"{u}rich, Photonics Laboratory, 8093 Z\"{u}rich, Switzerland }

\author{  Lukas Novotny }
\affiliation{ ETH Z\"{u}rich, Photonics Laboratory, 8093 Z\"{u}rich, Switzerland }

\author{  Paola Ayala }
\affiliation{ University of Vienna, Faculty of Physics, 1090 Wien, Austria }
\affiliation{ Yachay Tech University, School of Physical Sciences and  Nanotechnology, 100119-Urcuqu\'{\i} Ecuador }

\author{  Thomas Pichler }
\affiliation{ University of Vienna, Faculty of Physics, 1090 Wien, Austria }

\author{  Angel Rubio}
\affiliation{ Nano-Bio Spectroscopy Group and European Theoretical Spectroscopy Facility (ETSF), Universidad del Pa\'{i}s Vasco, CFM CSIC-UPV/EHU-MPC \& DIPC, 20018 San Sebasti\'{a}n, Spain }
\affiliation{ Max Planck Institute for the Structure and Dynamics of Matter, Hamburg, Germany }

\begin{abstract}
  Recently, the novel system of linear carbon chains inside of double-walled carbon nanotubes has extended the length of $sp^1$ hybridized carbon chains from 44 to thousands of atoms [L. Shi \emph{et al.}, Nat. Mater. \textbf{15}, 634 (2016)]. The optoelectronic properties of these ultra-long chains are poorly described by current theoretical models, which are based on short chain experimental data and assume a constant environment. As such, a physical understanding of the system in terms of charge transfer and van der Waals interactions is widely missing. We provide a reference for the intrinsic Raman frequency of polyynes \emph{in vacuo} and explicitly describe the interactions between polyynes and carbon nanotubes. We find that van der Waals interactions strongly shift the Raman frequency, which has been neither expected nor addressed before. As a consequence of charge transfer from the tube to the chain, the Raman response of long chains is qualitatively different from the known phonon dispersion of polymers close to the $\Gamma$-point. Based on these findings we show how to correctly interpret the Raman data, considering the nanotube's properties. This is essential for its use as an analytical tool to optimize the growth process for future applications.
\end{abstract}


\maketitle

Carbyne, the infinite $sp^1$ hybridized linear carbon chain, is often characterized by an intense Raman band associated with the eigenmode of the longitudinal-optical (LO) branch of the conjugated polymer close to the $\Gamma$-point (called \reflectbox{R}-mode in Zerbi's effective conjugation coordinate theory \cite{Castiglioni1988, Tian1990}). This $\Gamma$-mode vibration is characterized by an in-phase stretching of the chain's triple bonds. The current theoretical understanding of this material is founded on experimental data from colloidal $sp^1$-carbon oligomers that are sterically stabilized by bulky end-groups. The longest chains of this type to date contain only 22 triple bonds ($N$=44 carbon atoms) \cite{Chalifoux2010}. The development of a novel system, long linear carbon chains stabilized inside double-walled carbon nanotubes (LLCCs@DWCNTs), has allowed for the synthesis of chains containing at least 8,000 carbon atoms \cite{Shi2016}. These chains, which are long enough for proposed applications in composite materials \cite{Liu2013} and electronics \cite{Baughman2006}, are poorly described by current theory.

\begin{figure}
  \includegraphics[width=8.5cm,keepaspectratio=true]{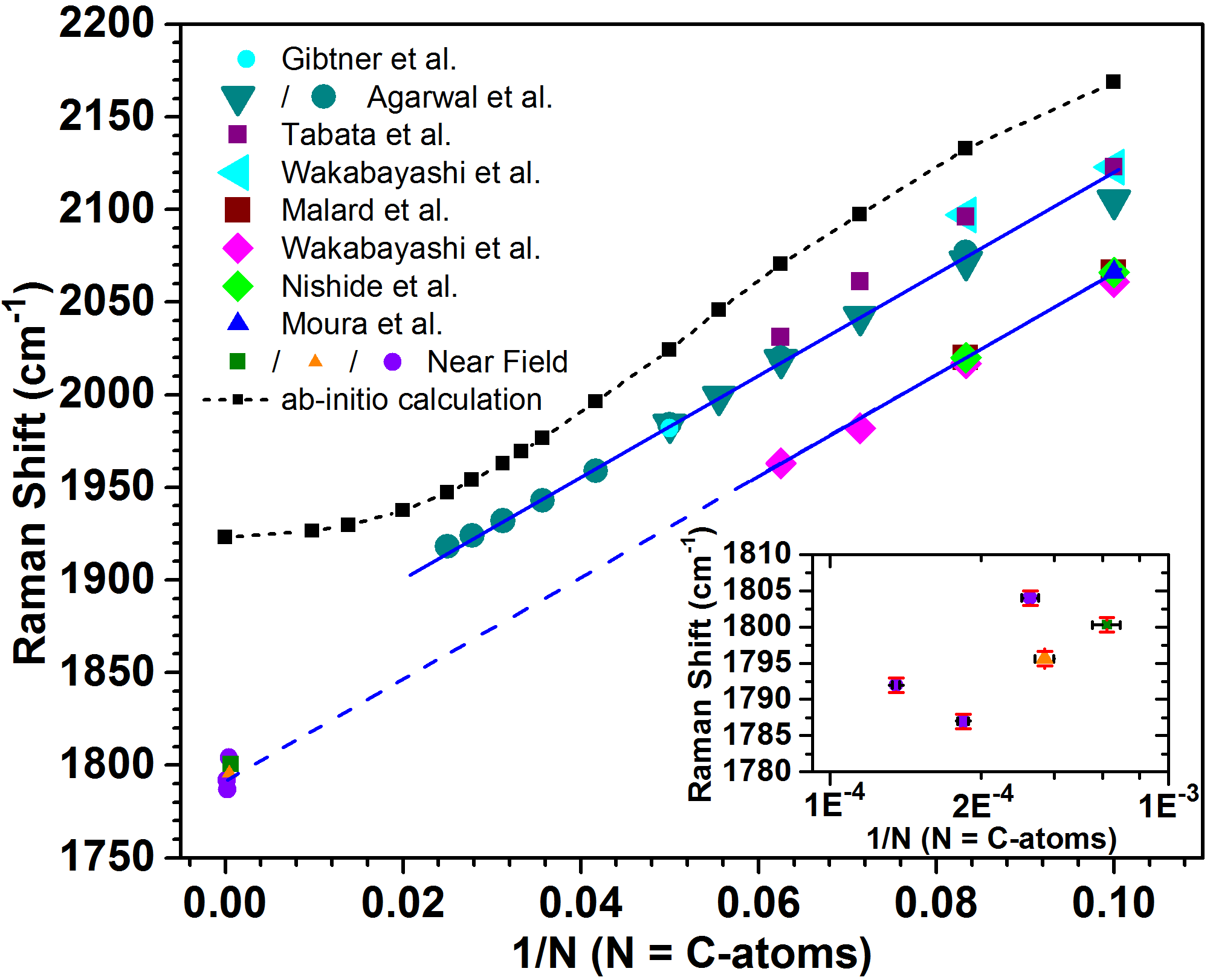}
  \caption{Raman response of polyynes as a function of inverse length, given by the number $N$ of carbon atoms. The solid lines are linear fits to the available data on chains with assigned lengths (upper solid line: colloidal chains \cite{Gibtner2002, Tabata2006, Wakabayashi2007, Agarwal2013}, lower solid line: chains inside CNTs \cite{Nishide2007,Wakabayashi2009, Moura2011}, dashed blue line: extrapolation to infinite chains). The black dashed line shows theoretical SCS-MP2 results \cite{Shi2016}. The inset shows data derived from near-field Raman microscopy images \cite{Shi2016}.\label{fig:5e}}
\end{figure}

The electronic and vibrational properties of carbyne are greatly influenced by the local environment as has been observed for short polyyne chains in solution~\cite{Gibtner2002, Tabata2006, Wakabayashi2007, Agarwal2013}, in CNTs~\cite{Nishide2007,Wakabayashi2009, Moura2011}, and on Ag surfaces~\cite{Tabata2006}. Theoretical models for short colloidal chains show a softening of the Raman frequency that saturates at around 1,900~\cm{} \cite{Yang2007c,Milani2008}; however, Raman frequencies below 1,800~\cm{} have been observed for chains in double- and multi-wall CNTs~\cite{Cupolillo2008, Castriota2008, Zhao2011a}.
We have now used scanning near-field optical spectroscopy to directly measure chain lengths and chain-band frequencies around 1,800~\cm{} for individual polyyne molecules of more than 1,000 atoms encapsulated in DWCNTs (inset in Fig.~\ref{fig:5e}). The measured distribution of data points illustrates two basic limitations of scaling schemes and current empirical models. First, the empirical parameters are supposed to correct the unknown error of the DFT force field and account for unknown environmental interactions that are not included in the physical model. Second, they assume a specific and constant environment that does not vary with chain length. This may be valid in colloids but cannot represent a CNT environment with a distribution of tube chiralities that interact differently with the chain, depending on their physical diameter and electronic properties.

Previous experiments on the interaction of CNTs with short chains of assigned length ($N$=8--12) have attributed the relative Raman shift, compared to colloidal chains, to CT from the tube to the chain \cite{Moura2009}. However, this conclusion was based on changes in the outer nanotube's G-band red shift \cite{Moura2011}, which can not be explained by a small CT alone, as the lifting of the Kohn anomaly by CT always leads a blue shift. Hence, this shift is most likely related to an internal strain inside the filled DWCNTs and not to a CT to the outer tubes.

Before Raman spectra of LLCCs@DWCNTs can be interpreted correctly and used for structural characterization, the following open questions must be answered: (1) To what extent are the LLCC@DWCNT spectra shifted by their local environment? (2) How does the variation in the measured frequencies depend on the nanotube chirality? (3) How does a specific medium affect the frequency softening for intermediate chain lengths ($N$=50-100 atoms), where \emph{ab initio} calculations \cite{Shi2016} show a saturation (black curve in Fig.~\ref{fig:5e}).

This work has two primary goals: (1) To obtain a reliable first-principles estimate of the Raman frequency of carbyne \emph{in vacuo} using highly accurate coupled-cluster CCSD(T) calculations, which will provide a reference point for any environmental effects. (2) To illuminate the quantum nature of the tube--chain interaction. This is done in two parts by evaluating the effect of the van der Waals (vdW) interaction on the chain's geometry and vibrational properties using an explicitly correlated method and the effect of charge transfer (CT) between CNTs and carbon chains in the short- and infinite-chain limit using hybrid density functional theory. Calculations were performed with the turbomole~\cite{tm7-0}, ORCA~\cite{ORCA3}, and VASP~\cite{Kresse1999} software, all details are given in the \supp. We find a surprisingly large and strongly length-dependent shift of the allowed Raman frequency of up to 280~\cm, which cannot be accounted for by CT alone. We show for the first time that vdW forces, which help to stabilize the chains inside the tube, modify the chain's bond-length alternation (BLA). Due to this quantum effect, the polarizability of the environment directly affects the vibrational and optical properties of the chain in the combined system.

\textbf{\emph{Polyynes in vacuo}}. To date there are no experimental IR or Raman spectra available for carbon chains \emph{in vacuo}. Therefore, only calculated spectra can currently serve as a reference point to quantify the environment effect in measured spectra.
The accurate \emph{ab initio} prediction of the BLA and $\Gamma$-mode frequency of polyynes is challenging and previous calculations are indeed inconsistent, ranging from cumulene to strongly alternated carbyne. The disagreement of \emph{ab initio} methods for long chains is well understood and can be attributed to the deficiencies of (semi-)local density functionals or to the incomplete description of local (dynamic) and long-range (static) correlation and has been studied extensively for polyenes \cite{Koerzdoerfer2012,Jacquemin2011,Wykes2015}. 

\begin{figure}
  \includegraphics[width=8.5cm,keepaspectratio=true]{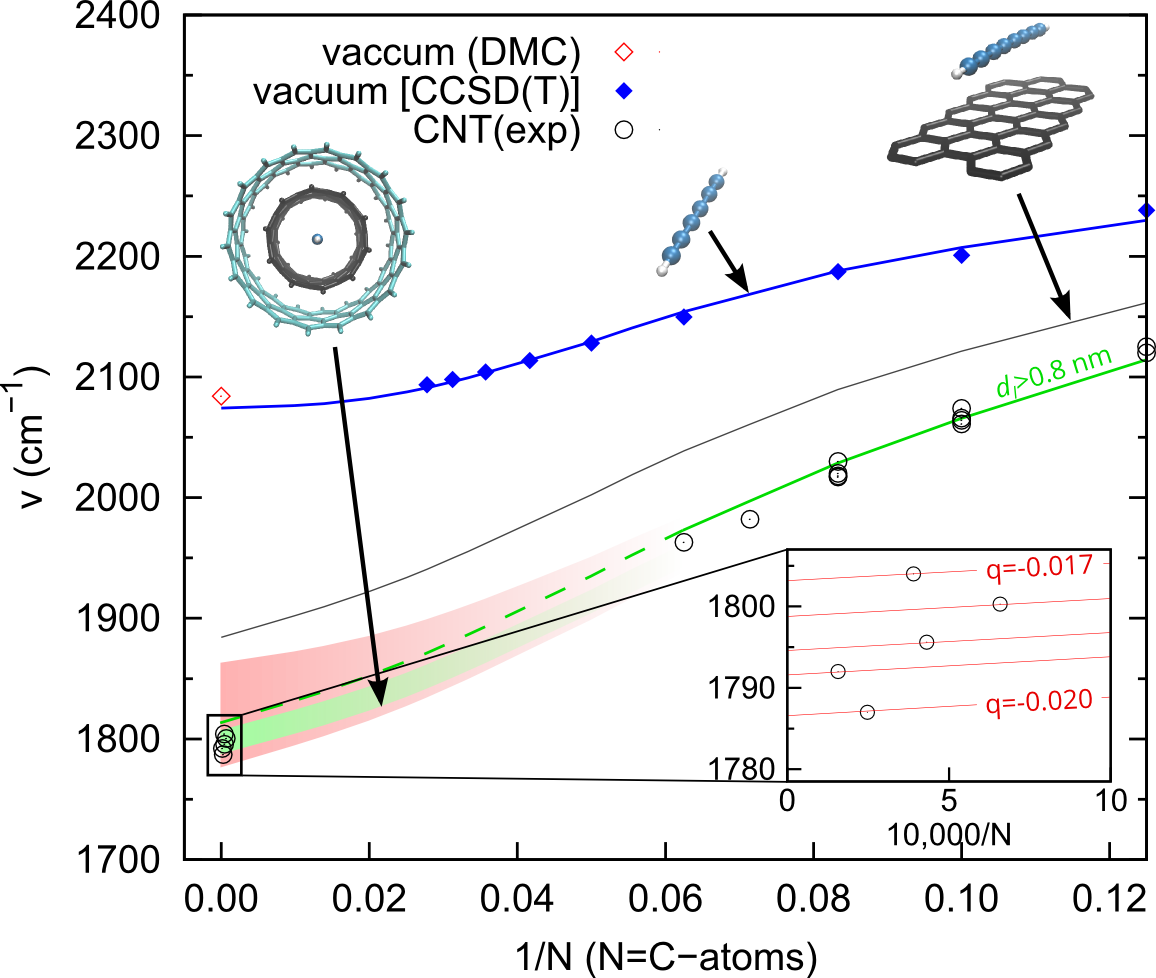}
  \caption{$\Gamma$-mode frequency of C$_N$H$_2$ in different environments as a function of inverse chain length.
  The green area shows the effect of vdW interactions varying with the inner tube diameter $d_I$ of a DWCNT within the range appropriate to host chains (0.65-0.75~nm). The solid/dashed green line represents the vdW interaction for larger $d_I$ and is the closest match to the experimental data of short chains. The upper limit is the interaction with graphene ($d_I$=$\infty$, grey line). The red area shows the effect of a variable CT ranging from 0.010 to 0.022~$e$ per chain atom (infinite chain) and decreasing with decreasing chain length. The inset shows the amount of charge transfer ($e$ per chain atom) required to reproduce the near-field Raman data, assuming the vdW shift associated with a constant $d_I$. Experimental data from refs \cite{Nishide2007,Wakabayashi2009, Moura2011, Shi2016}; DMC data from Mostaani \emph{et al.}~\cite{Mostaani2016}. \label{fig:chainlength}}
\end{figure}

Arguably the most precise methods that avoid these deficiencies and are computationally feasible are the coupled-cluster CCSD(T) and the diffusion Monte Carlo (DMC) methods. The quality of the latter depends on the careful choice of various parameters. To obtain the most unbiased results, we performed CCSD(T) calculations to obtain reference geometries and vibrational frequencies for C$_N$H$_2$ with $N$ ranging from 8 to 36. Our geometries are consistent with those of previous CCSD(T) calculations \cite{Zeinalipour-Yazdi2008} and our extrapolated BLA for the infinite chain is 0.125~\AA, slightly less than obtained from DMC \cite{Mostaani2016} (see Fig.~S1 in the \supp), and follows the same trend as polyenes~\cite{Barborini2015}. Longer chains are computationally unfeasible at the same level of theory and therefore we searched for the quantum chemical method that best reproduces the CCSD(T) data without scaling the force constants. Spin-component-scaled (SCS) variants of MP2 \cite{Gerenkamp2004} perform significantly better than any of the tested density functionals (see Table~S1 in the \supp). The calculated SCS-MP2 frequencies are very accurate for short chains, but the slope is too large (Fig.~S2). In order to extrapolate the CCSD(T) data to infinite chains, we scaled the SCS-MP2 data to fit the CCSD(T) points (blue line in Fig.~\ref{fig:chainlength}). This yielded a $\Gamma$-point frequency for carbyne \emph{in vacuo} of 2,075~\cm, very close to the DMC result by Mostaani \emph{et al.}~\cite{Mostaani2016} (red diamond in Fig.~\ref{fig:chainlength}).

In Fig.~\ref{fig:chainlength} we compare the $\Gamma$-mode frequencies of chains \emph{in vacuo}, as obtained from CCSD(T) theory, with experimental data from chains in CNTs. For the shortest chain, the $\Gamma$-mode frequency measured in a SWCNT is 118~\cm{} lower than the gas-phase reference calculation and the chain-length dispersion is different. The frequency shift continuously increases to 187~\cm{} for $N$=16 and reaches remarkable 270-290~\cm{} for LLCCs@DWCNTs.

\textbf{\emph{Chain length--tube chirality correlations}}. Assuming the growth conditions of LLCCs@DWCNTs allow the system to reach a thermal equilibrium distribution of chain lengths, it is clear that the average chain length must be different for each tube chirality as the tube--chain interaction energy depends on the physical inner tube diameter $d_I$ \cite{Shi2016}. This is corroborated by the sub-peak structure of the LLCC Raman band \cite{Zhao2003,Cupolillo2008,Shi2016}. Moreover, the pressure-dependency of the sub-peaks \cite{FerreiraAndrade2015} shows a direct dependence of the Raman shift on $d_I$, in addition to the shift from the nanotube’s electronic properties. Hence, the measured Raman frequencies cannot be described by one continuous $\nu(N)$ curve but rather a set of curves, each representing a specific tube diameter and chirality. This is illustrated in Fig.~\ref{fig:chainlength}, where we use a simple empirical formula to model the $N$-dependent environment shifts due to vdW interactions and CT, based on experimental data and our calculations, as will be discussed below.

\textbf{\emph{The effect of vdW interactions}}. The contribution of the dispersive or vdW force to the tube--chain interaction energy can be calculated with simple atom-pair additive potentials. Calculating the direct effect of vdW interactions on the geometric and vibrational properties, in contrast, is computationally difficult due to its many-body nature and therefore generally neglected. Additionally, experiments cannot easily distinguish it from competing physical effects such as CT and static polarization. Tentatively, vdW interactions have been associated with redshifts of the resonance energy of short chains inside SWCNTs \cite{Moura2009, Moura2011}. The observed trend, increasing redshift with CNT diameter, is surprising considering that the calculated interaction energy as a function of tube radius is largest in magnitude close to the vdW distance (3.3~\AA{}) and decreases for larger diameters \cite{Shi2016}. Moreover, when the tube diameter is reduced in high-pressure experiments, a redshift in the Raman frequency of the chain is observed \cite{FerreiraAndrade2015}.

Theoretically, it is clear that the parameters relevant for the vdW interaction, such as chain polarizability ($\alpha$) and ionization potential (IP), depend on the chain's BLA (see Fig.~\ref{fig:polarizability}); however, it is difficult to develop an empirical model based on such properties because the tube--chain separation is much smaller than the molecular extension and it is unclear how effective atomic parameters can be defined that reflect the locality of the vdW interaction. \emph{Ab initio} calculations that quantify the effect are challenging as the relevant electron correlation must be treated explicitly, therefore limiting the size of potential model systems.

\begin{figure}
  \includegraphics[width=8.5cm,keepaspectratio=true]{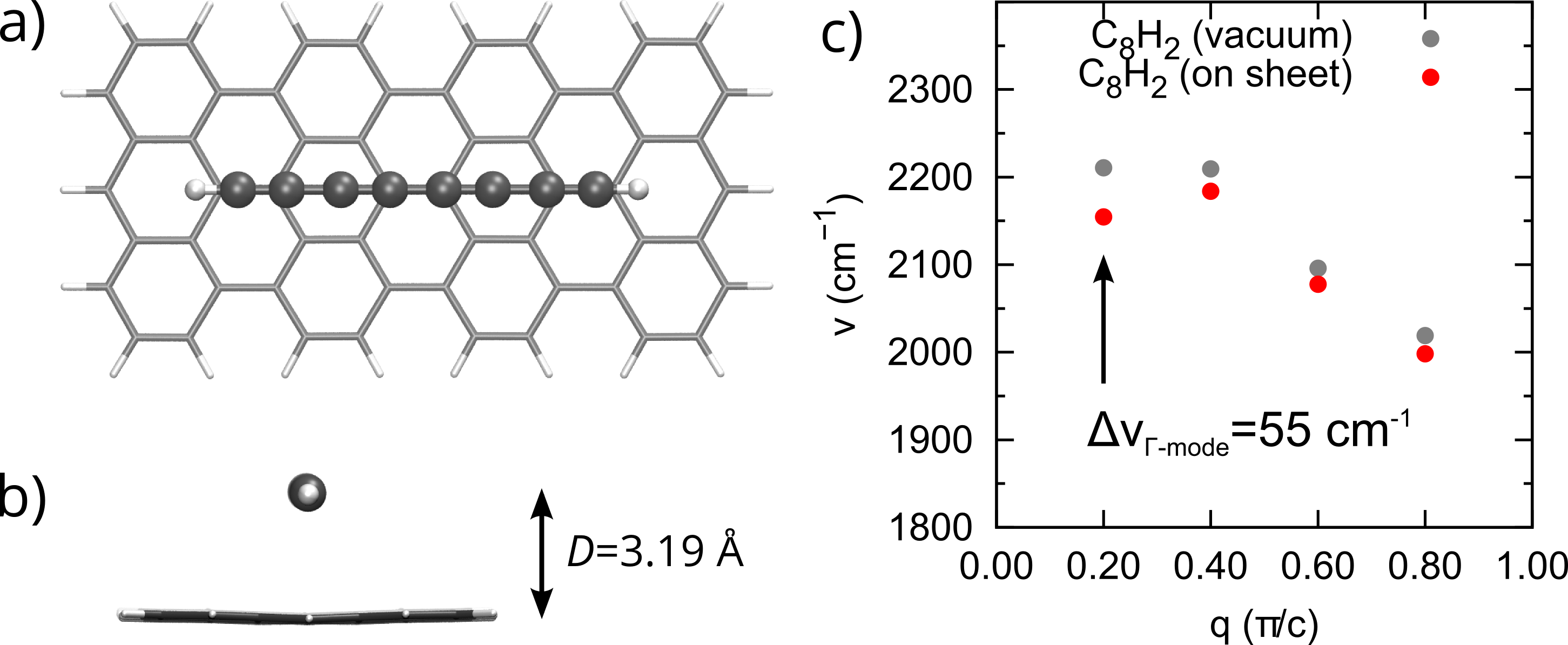}
  \caption{Adsorption of a C$_8$H$_2$ chain on a graphene sheet. (a) Top view, (b) side view of the relaxed (SCS-MP2) geometry. (c) LO-branch vibrational frequencies.\label{fig:armchair-c8h2}}
\end{figure}

As a minimal model that mimics a semiconducting tube of infinite diameter, we placed C$_{8}$H$_2$ on a hydrogen-terminated graphene sheet (Fig.~\ref{fig:armchair-c8h2}) and calculated the fully relaxed geometry and vibrational spectrum. Indeed, SCS-MP2 predicts a reduced BLA and softening of the $\Gamma$-mode frequency by 55~\cm{} with respect to the isolated chain (Table S2 and Fig.~\ref{fig:armchair-c8h2}). Comparative DFT-D calculations (Table~S2), which reproduce the correct chain--sheet separation while neglecting the vdW correlation, show that other effects, in particular CT, are small and do not significantly shift the vibrational frequency. Therefore, the length-dependent frequency shift due to the vdW interaction can be obtained as the difference between the total environment shift and the shift due to CT (\emph{vide supra}). LLCCs are expected to exist in similar narrow inner tubes ($d_I$=0.65--0.75~nm), which provide a maximal vdW interaction. The resulting frequency span is shown in Fig.~\ref{fig:chainlength} as green area.

\textbf{\emph{The role of charge transfer}}.
Previous theoretical studies predicted a CT between carbyne and host tubes that reduces the BLA~\cite{Rusznyak2005,Tapia2010}. To test whether CT is also relevant for short chains and can explain the observed red shift as compared to colloidal chains, we performed DFT calculations of C$_{10}$H$_2$ in different SWCNTs (Table~\ref{tab:tubes}).
We find only a small redshift (21--25~\cm) for the $\Gamma$-mode in different tubes, consistent with a very small calculated CT from the tube to the chain.

The small redshift is not surprising as we find no indication of hybridization between the chain and tube levels and the chain LUMO remains empty. Further, the effect of CT on the frequency may be overestimated by the PBE functional due to the above mentioned error in the phonon dispersion and slope of $\nu(N)$. This means that the redshift between our CCSD(T) calculations (2,201~\cm{} \emph{in vacuo}) and experimental data (2,061~\cm{} in SWCNT~\cite{Wakabayashi2009}) for C$_{10}$H$_2$ can not be accounted for by CT alone.

\begin{table}
\caption{~C$_{10}$H$_2$ \emph{in vacuo} and in different CNTs.$^a$\label{tab:tubes}}
\begin{ruledtabular}
\begin{tabular}{lcccrr}
 & $d_I$ & BLA & $\nu$ & $\Delta\nu$ & $Q_{\rm chain}$ \\
\hline
vacuum &  & 0.1040 & 2133 & 0 & 0.000 \\
C$_{10}$H$_2$@(12,0) & 0.94 & 0.1038 & 2112 & -21 & -0.026 \\
C$_{10}$H$_2$@(6,6) & 0.81 & 0.1034 & 2110 & -23 & -0.044 \\
C$_{10}$H$_2$@(10,0) & 0.78 & 0.1035 & 2109 & -24 & -0.047 \\
C$_{10}$H$_2$@(10,0)@(18,0) & 0.78 & 0.1035 & 2108 & -25 & -0.056 \\
\end{tabular}
\end{ruledtabular}
  \\\footnotesize{$^a$PBE calculations of large supercells [6-fold for (m,0) and 9-fold for (6,6) tubes, see \comp{} for details]. Columns 2--6 show the (inner) tube diameter $d_I$ (pm), BLA (\AA{}), $\Gamma$-mode frequency (\cm), shift with respect to vacuum (\cm), and net charge of the chain ($e$).}
\end{table}

\begin{figure}
  \includegraphics[width=7.5cm,keepaspectratio=true]{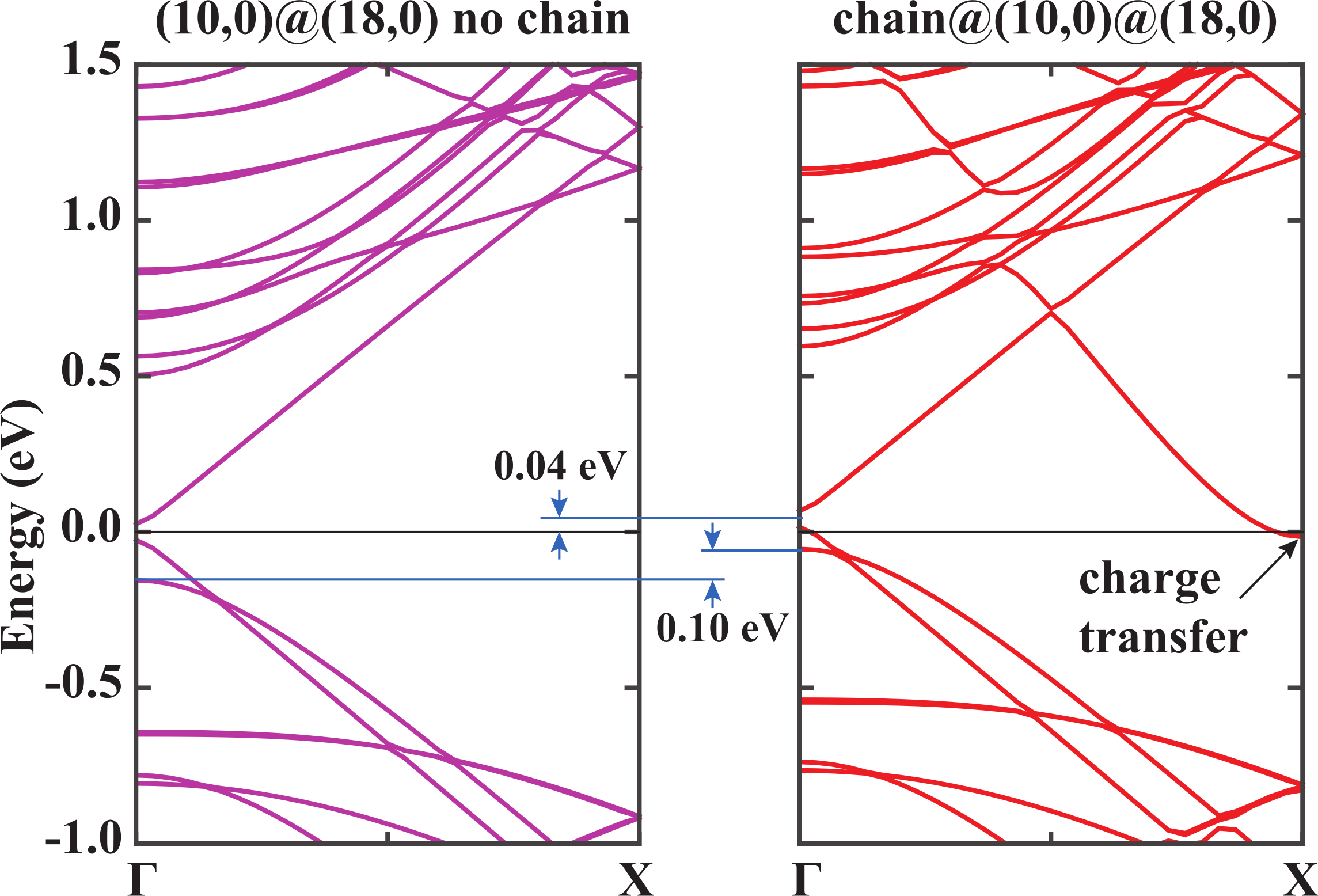}
  \caption{Band structures of a (10,0)@(18,0) DWCNT empty (left) and encapsulating an infinite polyyne chain (right).\label{fig:bandstructure}}
\end{figure}

\begin{figure}
  \includegraphics[width=4.25cm,keepaspectratio=true]{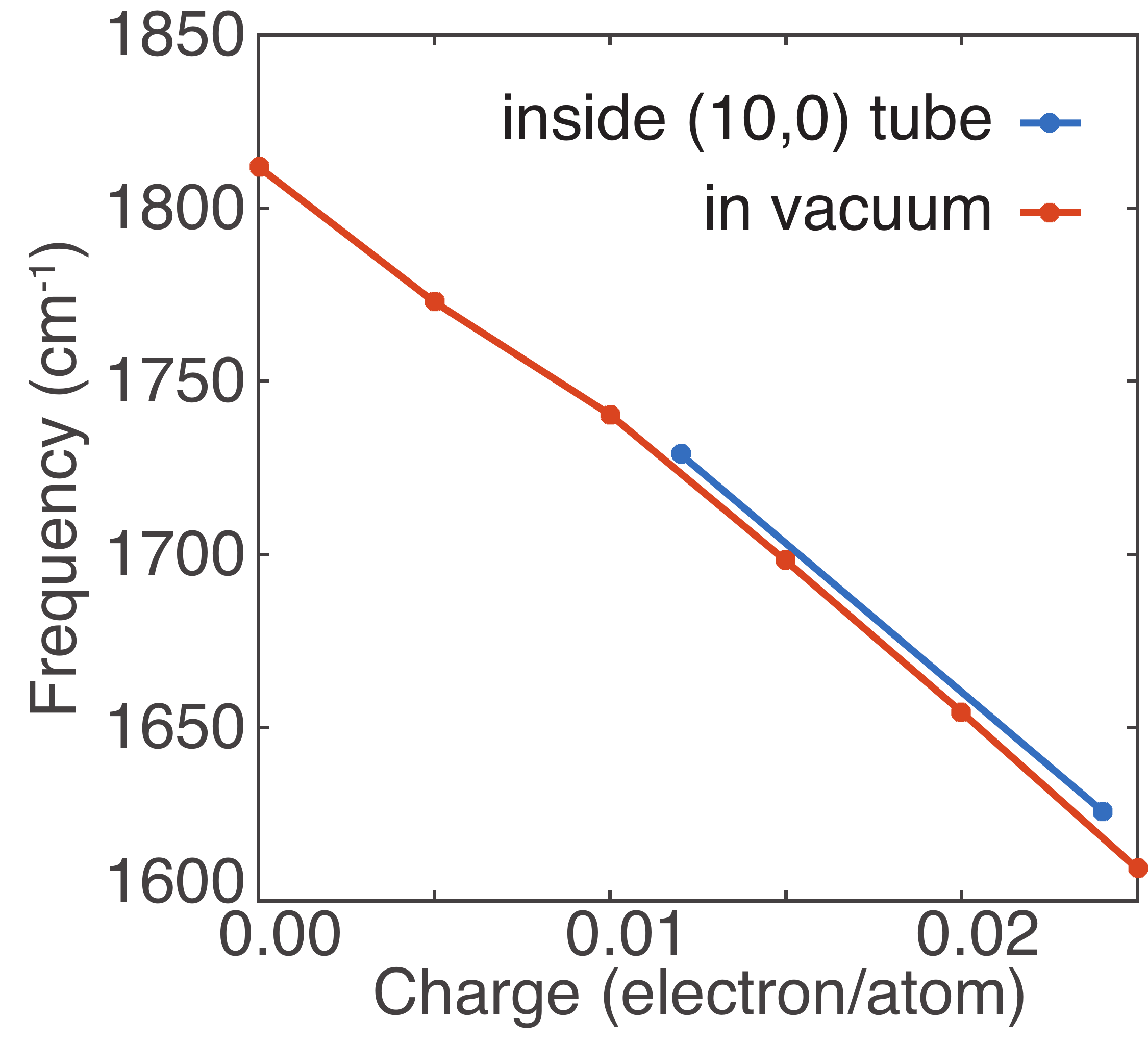}
  \includegraphics[width=4.25cm,keepaspectratio=true]{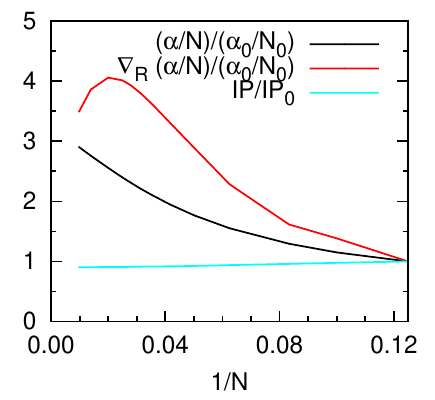}
  \caption{Left: HSE calculation of the $\Gamma$-mode frequency of the infinite chain as a function of excess charge per chain atom.\label{fig:charge-frequency} Right: Polarizability per C-atom, its derivative with respect to the displacement $R$ along the $\Gamma$-mode, and IP. Properties are normalized with respect to C$_8$H$_2$ ($\alpha_0$, $N_0$, and IP$_0$).\label{fig:polarizability}}
\end{figure}

For long chains in double- or multi-wall CNTs the chain LUMO approaches the typical work function (WF) of these CNTs (Fig.~S3), suggesting that the CT can be larger. We performed a DFT (PBE functional) supercell calculation of the infinite polyyne chain (HSE vacuum geometry) inside a (10,0)@(18,0) DWCNT. Fig.~\ref{fig:bandstructure} compares the band structures of the pristine and filled DWCNT. The chain LUMO band overlaps with the valence band of the metallic outer tube, which leads to a CT to the chain of 0.022~$e$ per chain-atom. This CT is about one order of magnitude larger than that obtained for short chains and is comparable to the calculated inter-wall CT of DWCNTs \cite{Zolyomi2008}.

The applied GGA functional, however, does not properly describe the BLA and phonon dispersion of the infinite chain. We therefore did not calculate these properties inside the DWCNT, but used the obtained CT as a parameter to estimate the effect on geometry and phonons in a smaller model system using more accurate methods. Fig.~\ref{fig:charge-frequency} shows the Raman frequencies of carbyne \emph{in vacuo} and inside a (10,0) SWCNT as a functional of electron doping, which were calculated with the HSE hybrid functional and relaxed chain geometry. Inside the SWCNT, the excess charge localizes on the chain. Therefore, both calculations show nearly the same BLA and $\Gamma$-mode frequency for a given amount of charge. In agreement with earlier LDA calculations \cite{Rusznyak2005}, the relation is clearly linear and the frequency redshifts by 176~\cm{} for a CT of 0.022~$e$ per chain atom.

As the HSE functional approximates PBE0, which overestimates the phonon dispersion (\emph{vide supra}), we also performed SCS-MP2 calculations of the charged infinite chain (circular boundary conditions, see \comp). We obtained a frequency shift of 134~\cm{} for a CT of 0.022~$e$ per chain atom. For both the hybrid DFT and the SCS-MP2 result, the frequency shift due to CT is clearly smaller than the total environment shift of 280~\cm{}. A major contribution must therefore be arise from vdW interactions.

Even for a constant inner tube diameter and its associated vdW shift, the CT can vary with the electronic structure of the outer tube. The red area in Fig.~\ref{fig:chainlength} illustrates the frequency span resulting from a CT of 0.010-0.022~$e$ per chain atom. The inset shows the amount of CT required to reproduce the experimental frequencies assuming a constant vdW interaction. Considering that the vdW interaction will increase with decreasing inner tube diameter, this range of CT can be seen as an upper limit. Raman data from short chains in CNTs is associated with larger tube diameters ($d_I$>0.9~nm) and the gap between the tube's WF and the chain's LUMO is large. Hence, the frequency shifts in this range depend primarily on the chain length and the experimental data can be fitted by a single line in Fig.~\ref{fig:chainlength}.

In summary, our results show that long encapsulated chains interact with the nanotube in a way that qualitatively changes the Raman response and its dependence on the inverse chain length. The saturating behavior known from the phonon dispersion of polymers close to the $\Gamma$-point does not apply to these systems, which is a consequence of a variable charge transfer and an unexpectedly strong effect of the van der Waals interaction. We have for the first time disentangled these two effects, which strongly depend on the chain length and together shift the chain's Raman band by 118-290~\cm{}. The huge van der Waals shift, required to reproduce the Raman data of both short and long carbon chains, is also interesting from a theoretical point because currently no approaches exists that can describe it correctly for any realistic system.
For ultra-long chains, i.e., confined carbyne, we predict an increased CT from the tube to the chain, which depends on the CNT's electronic properties.
For the longest measured chains we estimate an upper limit of the CT of 0.02~$e$ per chain atom, which varies by 15\% depending on the tube chirality and accounts for about 40\% of the shift. For short chains measured in wider CNTs the CT is smaller and causes shifts of 6-25~\cm.
Considering that both CT and vdW interactions depend on the tube chirality and that the equilibrium distribution of chain lengths is correlated with these tube properties, the limits of previous simple scaling schemes become obvious. On the other hand, our results show the pathway how to correctly assign the chain length from Raman spectroscopy. If additional information about the inner tube diameter and the electronic properties of the nanotube are considered, the chain's Raman band can be used as an analytic tool to optimize the growth of confined carbyne, which we see as an essential step towards accessing their theoretically outstanding application potential.

\begin{acknowledgments}
  M.W. and A.R. acknowledge financial support by the European Research Council (ERC-2010-AdG-267374), Spanish MINECO (FIS2013-46159-C3-1-P), Grupos Consolidados (IT578-13), AFOSR Grant No. FA2386-15-1-0006 AOARD 144088, H2020-NMP-2014 project MOSTOPHOS (GA no. 646259), and COST Action MP1306 (EUSpec).
  Technical and human support provided by IZO-SGI (SGIker) of UPV/EHU.
  S.C. acknowledges financial support from the Marie Curie Grant FP7-PEOPLE-2013-IEF project ID 628876. T.P. acknowledges support from the Austrian Science Fund (FWF, NanoBlends I 943–N19). L.S. acknowledges the scholarship supported by the China Scholarship Council. Z.J.L. and L.N. acknowledge Swiss National Science Foundation (CR2212-152944).
\end{acknowledgments}

\bibliography{manuscript}

\clearpage

\widetext
\begin{center}
\textbf{\large Supplemental Material for:\\ Polyyne Electronic and Vibrational Properties under Environmental Interactions}
\end{center}

\setcounter{equation}{0}
\setcounter{figure}{0}
\setcounter{table}{0}
\setcounter{page}{1}
\makeatletter
\renewcommand{\theequation}{S\arabic{equation}}
\renewcommand{\thefigure}{S\arabic{figure}}
\renewcommand{\thetable}{S\arabic{table}}
\renewcommand{\thepage}{S\arabic{page}}

\section{Computational Details}

Hybrid DFT (PBE0 functional [S1]) and SCS-MP2 calculations were performed with the turbomole~[S2] software (versions 6.1, 6.6, and 7.0). Finite C$_N$H$_2$ chains with $N$ up to 102 were fully optimized at the SCS-MP2 level with turbomole, using a cc-pVDZ basis set. For $N$ up to 40, the frequency of the resonant Raman active mode was calculated numerically. For $N$>40, we used the PBE0 eigenmodes to displace the SCS-MP2 geometry and generate a harmonic fit of the potential energy surface of SCS-MP2. This approach introduces an error of less than 1~\cm~(for $N$>20) in the resulting frequency. The SCS-MP2 frequency for the infinite chain was obtained by a series of calculations of carbon rings (circular boundary condition). The ring size was increased up to 144 carbon atoms to achieve converged bond lengths and frequencies.

For the frequency calculations of the \emph{finite} C$_{10}$H$_2$ chain in vacuo and inside SWCNT and DWCNT, we used the PBE and HSE functionals as implemented in the VASP 5.3.5 code [S3]. For the HSE functional, a range-separation parameter HFSCREEN=0.2 was used. Like in the SCS-MP2 calculations, we employed the eigenmodes of gas-phase PBE0 calculations to obtain the force constants from the gradient of the displaced minimum-energy geometry. The following supercells and k-point samplings were used: 6-fold supercell of (10,0), (12,0), and (10,0)@(18,0) with 1 k-point and 9-fold supercell of (6,6) with 1 k-point.

For the PBE band-structure calculations of the \emph{infinite chain} inside a (10,0)@(18,0) DWCNT, we used a 3-fold supercell of the DWCNT, which hosts 10 chain atoms. We used the equilibrium bond lengths of the chain in vacuo as obtained with the HSE functional. We did not optimize the chain inside the tube, because the GGA functional does not properly describe the BLA and favours a cumulenic structure inside CNTs [S4]. The box size was adjusted to the chain geometry. 31 irreducible k-points were used.

For the C$_8$H$_2$@graphene calculations, a saturated graphene sheet of 78 atoms was used (Fig.~5 in the main article). The geometry was fully relaxed with turbomole using SCS-MP2 (cc-pVDZ basis set), or DFT [SV(P) basis set]. The eigenmodes from gas-phase PBE0 calculations were employed to obtain the force constant of the $\Gamma$-mode. For C$_8$H$_2$ in vacuo, the resulting errors were smaller than 1~\cm~when compared with the frequency from the full analytical Hessian.

CCSD(T)/cc-pVDZ geometry optimizations were performed with turbomole version 7.0.2 using symmetry-adapted numerical gradients. CCSD(T) frequencies were obtained using the displacements of the SCS-MP2 eigenmodes or (for $N$=20) those of the PBE0 Hessian (see above).

Benchmark calculations (Table~S1) for the geometry of C$_{12}$H$_2$ were performed with ORCA using the cc-pVDZ basis set, except for the functionals LC-PBE, M11, $\omega$B97XD, BMK, HISS, tHCTh, and CAM-B3LYP, which were performed with Gaussian~[S5].

\clearpage

\section{Geometry Benchmark of C$_N$H$_2$}

\begin{figure}[!h]
  \includegraphics[width=12cm,keepaspectratio=true]{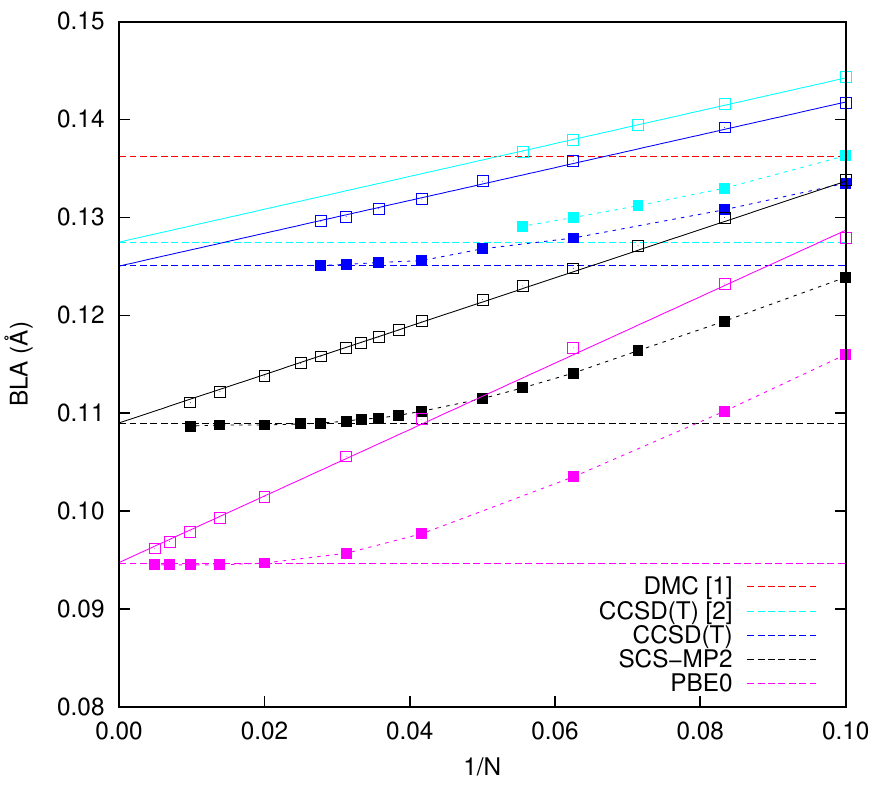}
  \caption{Bond-length alternation (BLA) as a function of inverse chain length ($N$ is the number of C-atoms). Full symbols show the BLA at the chain center. Empty symbols show the average BLA over the chain (continuous line: linear fit). Horizontal lines mark the BLA of the infinite chain, which is extrapolated from the linear fit in case of CCSD(T). The validity of the linear extrapolation is confirmed by the SCS-MP2 and PBE0 BLA of the converged series (H-termination and circular boundary conditions yield the same value).\label{fig:bla}}
\end{figure}

\begin{table*}[!h]
\caption{~C$_{12}$H$_2$ RMSD of the energy gradient (atomic units) of the CCSD(T) geometry.\label{tab:armchair-c8h2}}
\begin{tabular}{ll}
\hline
Method & RMSD \\
\hline
CCSD(T) & 0.0000 \\
ES-MP2$^a$ & 0.0028 \\
ES-MP2$^b$ & 0.0033 \\
SCS-MP2 & 0.0042 \\
SCS(MI)-MP2 & 0.0045 \\
B2GP-PLYP & 0.0052 \\
mPW2PLYP & 0.0053 \\
B3LYP & 0.0065 \\
M06 & 0.0067 \\
PBE0 & 0.0068 \\
B2PLYP & 0.0071 \\
PW6B95 & 0.0076 \\
BMK & 0.0076 \\
TPSS0 & 0.0078 \\
HISS & 0.0119 \\
MP2 & 0.0136 \\
$\omega$B97XD & 0.0143 \\
M062X & 0.0146 \\
tHCTh & 0.0146 \\
CAM-B3LYP & 0.0150 \\
PBE & 0.0173 \\
BH\&HLYP & 0.0188 \\
M11 & 0.0192 \\
LC-PBE & 0.0249 \\
HF & 0.0431 \\
\hline
\end{tabular}
  \\\footnotesize{$^a$MP2 with equally-scaled spin components (ES-MP2), optimized scaling parameter of 0.763.\\
                  $^b$ES-MP2 with scaling parameter of 0.73.}
\end{table*}

\begin{figure}[!h]
  \includegraphics[width=8.5cm,keepaspectratio=true]{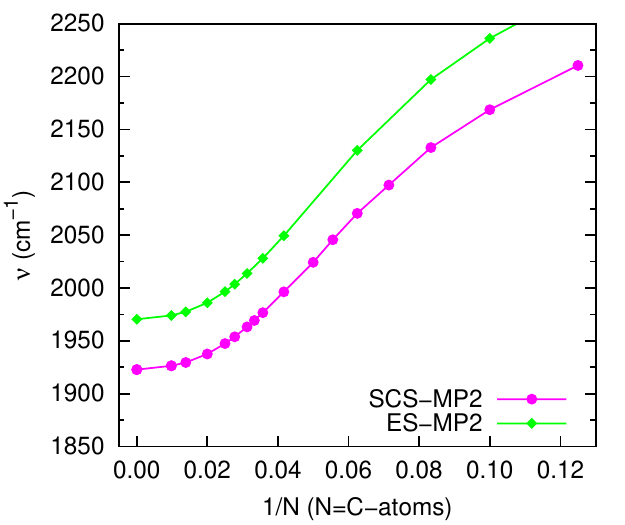}
  \caption{$\Gamma$-mode frequency of C$_N$H$_2$ as a function of inverse chain length. Comparison of different MP2 variants.\label{fig:chainlength-mp2}}
\end{figure}

\begin{figure}[!h]
  \includegraphics[width=8.5cm,keepaspectratio=true]{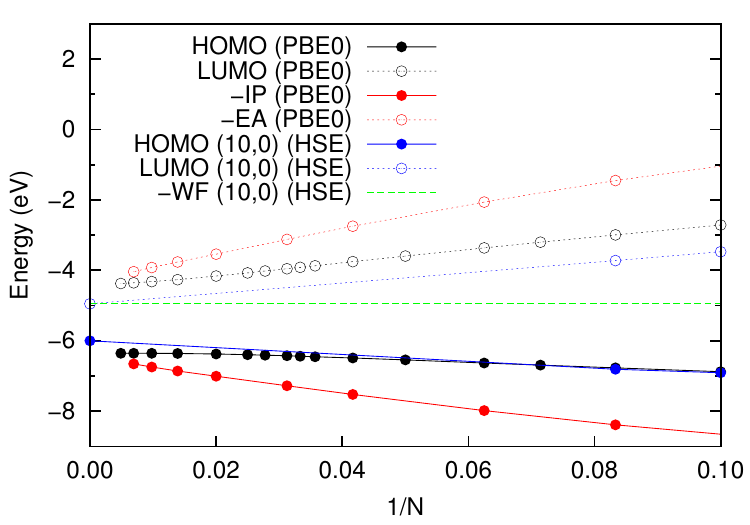}
  \caption{Comparison of electronic levels, EA, and IP of C$_N$H$_2$ in vacuo and inside SWCNT (HSE calculation) and the work function of a (10,0) SWCNT (HSE calculation).\label{fig:homolumo}}
\end{figure}  

\begin{table*}[!h]
\caption{~$\Gamma$-mode frequency (\cm) and BLA (\AA{}) of C$_8$H$_2$ in vacuo and adsorbed on a graphene sheet.\label{tab:armchair-c8h2}}
\begin{ruledtabular}
\begin{tabular}{lccrccrcc}
 method & $\nu_{\rm vacuo}$ & $\nu_{\rm sheet}$ & $\Delta\nu$ & BLA$_{\rm vacuo}$ & BLA$_{\rm sheet}$ & $\Delta$BLA & $D^a$ & $Q_{\rm chain}$ \\
\hline
SCS-MP2 & 2209.9 & 2154.8 & -55.1 & 0.1392 & 0.1298 & -0.0094 & 3.19 & -0.010 \\
PBE-D2 & 2182.3 & 2175.9 & -6.4 & 0.1101 & 0.1094 & -0.0007 & 3.19 & -0.020 \\
PBE-D3 & 2183.8 & 2180.1 & -3.7 & 0.1099 & 0.1095 & -0.0004 & 3.37 & -0.012 \\
\end{tabular}
\end{ruledtabular}
  \\\footnotesize{$^a$Average separation (\AA{}) of the chain molecules from the surface.}
\end{table*}

\clearpage
\begin{description}
  \item[{[S1]}] J. P. Perdew, M. Emzerhof, and K. Burke, J. Chem. Phys. \textbf{105}, 9982 (1996).
  \item[{[S2]}] \emph{TURBOMOLE   V7.0   2015,   a   development   of   University   of   Karlsruhe   and   Forschungszen-
trum   Karlsruhe   GmbH,   1989-2007,    TURBOMOLE   GmbH,   since   2007;    available   from}
\texttt{http://www.turbomole.com}.
  \item[{[S3]}] G. Kresse and D. Joubert, Phys. Rev. B \textbf{59}, 1758 (1999).
  \item[{[S4]}] \'{A}. Ruszny\'{a}k, V. Z\'{o}lyomi, J. K\"{u}rti, S. Yang, and M. Kertesz, Phys. Rev. B \textbf{72}, 155420 (2005).
  \item[{[S5]}] Gaussian 09, Revision D.01, M. J. Frisch, G. W. Trucks, H. B. Schlegel, G. E. Scuseria, M. A. Robb,
J. R. Cheeseman, G. Scalmani, V. Barone, B. Mennucci, G. A. Petersson, H. Nakatsuji, M. Caricato,
X. Li, H. P. Hratchian, A. F. Izmaylov, J. Bloino, G. Zheng, J. L. Sonnenberg, M. Hada, M. Ehara, K.
Toyota, R. Fukuda, J. Hasegawa, M. Ishida, T. Nakajima, Y. Honda, O. Kitao, H. Nakai, T. Vreven,
J. A. Montgomery, Jr., J. E. Peralta, F. Ogliaro, M. Bearpark, J. J. Heyd, E. Brothers, K. N. Kudin,
V. N. Staroverov, T. Keith, R. Kobayashi, J. Normand, K. Raghavachari, A. Rendell, J. C. Burant,
S. S. Iyengar, J. Tomasi, M. Cossi, N. Rega, J. M. Millam, M. Klene, J. E. Knox, J. B. Cross, V.
Bakken, C. Adamo, J. Jaramillo, R. Gomperts, R. E. Stratmann, O. Yazyev, A. J. Austin, R. Cammi,
C. Pomelli, J. W. Ochterski, R. L. Martin, K. Morokuma, V. G. Zakrzewski, G. A. Voth, P. Salvador,
J. J. Dannenberg, S. Dapprich, A. D. Daniels, O. Farkas, J. B. Foresman, J. V. Ortiz, J. Cioslowski,
and D. J. Fox, Gaussian, Inc., Wallingford CT, 2013.
\end{description}

\end{document}